\documentclass{pasj00}
\draft

\begin{document}
\SetRunningHead{Nakashima et al.}{2015/06/22}

\title{Methanol Observation of IRAS 19312+1950:\\ A Possible New Type of Class I Methanol Masers}

\author{Jun-ichi Nakashima,\altaffilmark{1,2}
Andrej M. Sobolev,\altaffilmark{3}
Svetlana V. Salii,\altaffilmark{1}
Yong Zhang,\altaffilmark{2,4}\\
Bosco H. K. Yung,\altaffilmark{2}
Shuji Deguchi,\altaffilmark{5}}
\altaffiltext{1}{Department of Astronomy and Geodesy, Institute of Natural Sciences, Ural Federal University,\\ Lenin Avenue 51, 620000, Ekaterinburg, Russia}
\altaffiltext{2}{Department of Physics, University of Hong Kong, Pokfulam Road, Hong Kong, China}
\altaffiltext{3}{Astronomical Observatory, Institute of Natural Sciences, Ural Federal University,\\ Lenin Avenue 51, 620000, Ekaterinburg, Russia}
\altaffiltext{4}{Space Astronomy Laboratory, Faculty of Science, The University of Hong Kong, Pokfulam Road, Hong Kong, China}
\altaffiltext{5}{Nobeyama Radio Observatory, National Astronomical Observatory of Japan, Minamimaki, Minamisaku, Nagano 384-1305, Japan}
\email{nakashima.junichi@gmail.com}

%


\maketitle


\begin{abstract}
We report the result of a systematic methanol observation toward IRAS 19312+1950. The properties of the SiO, H$_2$O and OH masers of this object are consistent with those of mass-losing evolved stars, but some other properties are difficult to explain in the standard scheme of stellar evolution in its late stage. Interestingly, a tentative detection of radio methanol lines was suggested toward this object by a previous observation. To date, there are no confirmed detections of methanol emission towards evolved stars, so investigation of this possible detection is important to better understand the circumstellar physical/chemical environment of IRAS 19312+1950. In this study, we systematically observed multiple methanol lines of IRAS~19312+1950 in the $\lambda=3$~mm, 7~mm, and 13~mm bands, and detected 6 lines including 4 thermal lines and 2 class I maser lines. We derived basic physical parameters including kinetic temperature and relative abundances by fitting a radiative transfer model. According to the derived excitation temperature and line profiles, a spherically expanding outflow lying at the center of the nebulosity is excluded from the possibilities for methanol emission regions. The detection of class I methanol maser emission suggests that a shock region is involved in the system of IRAS 19312+1950. If the central star of IRAS 19312+1950 is an evolved star as suggested in the past, the class I maser detected in the present observation is the first case detected in an interaction region between an evolved star outflow and ambient molecular gas. 
\end{abstract}


\section{Introduction}
SiO maser sources are, in most cases, identified as being associated with mass-losing evolved stars of either asymptotic giant branch (AGB) stars, post-AGB stars or red supergiants (RSGs). In fact, the maser properties of SiO, H$_2$O and OH masers associated with IRAS~19312+1950 (I19312, hereafter) are consistent with those of AGB/post-AGB stars or RSGs \citep{nak11}, and the properties of the central infrared point source are also consistent with those of mass-losing evolved stars \citep{mur07}. Nevertheless, some other characteristics of this object cannot be explained in the standard scheme of late-stage stellar evolution. For example, an extended nebulosity with point symmetric structure, which shows complicated knots, surrounds a spherically expanding molecular outflow lying at the center \citep{nak00,nak04a,deg04,nak05,mur07}, the gas-mass derived from CO lines is relatively high for an evolved star (10--15 M$_\odot$; \cite{deg04}), and a rich set of molecular species has been detected toward the nebulosity (so far 22 molecular species have been detected; \cite{deg04,nak05}).

Since SiO maser emission is occasionally detected toward dense cores in molecular clouds (for example, Ori IRc 2, W51 IRs 2, and Sgr B2 MD5; \cite{has86}), which are detectable in various molecular lines, one possible explanation might be that I19312 is a young stellar object (YSO) embedded in a molecular cloud. However, the characteristics of I19312 are clearly different from such SiO maser sources embedded in molecular clouds. For instance,  I19312 shows an isolated point-like feature in mid-infrared images (see, e.g., \cite{nak11}),
while all the SiO maser sources found in molecular clouds are embedded in a huge star forming region, of which the background is generally very bright at mid-infrared wavelengths. Additionally, there are no clear star forming activities in the vicinity of I19312, such as an enhancement of the number density of stars.

As of now, three hypotheses are being suggested as possible explanations for the origin of I19312 (see, e.g., \cite{nak11}): (1) a red nova formed by the merger of two main sequence stars (or two stars going to main sequence), (2) a mass-losing evolved star physically associated with a small, isolated molecular cloud, and (3) a mass-losing evolved star behind a small isolated molecular cloud, which is not physically associated with the star (but the star and molecular cloud shares the same systemic velocity). However, none of these hypotheses can adequately explain the observational characteristics of I19312.

Interestingly, \citet{deg04} reported a tentative detection of radio methanol lines, but this tentative detection is unexpected because the most likely interpretation of I19312 are that it is a mass-losing evolved star. In fact, there have been no previous secure detections of radio methanol lines toward any evolved stars, though some sensitive methanol searches have been made toward the samples of evolved stars (see, e.g., \cite{cha97,for04,he08}). 

If the methanol detection were to be confirmed it would be of great significance in studies of the physical and chemical environments of evolved stars. The detection of multiple methanol lines enables essential physical parameters of the emission region to be determined, which would be invaluable in understanding this unusual source.

In the present research, we systematically observed multiple methanol lines at $\lambda=3$~mm, 7~mm, and 13~mm bands toward I19312, and securely detected 6 methanol lines including 4 thermal lines and 2 class I maser lines. The detected class I masers seem not to be classified into any known categories of methanol masers. We obtained some basic physical parameters by constructing a simple radiative transfer model. Based on the results, we discuss the nature of the methanol emission region, implying possible locations of the emission region of methanol lines.


\section{Details of Observation and Results}
The observations were made with the Nobeyama Radio Observatory (NRO) 45~m telescope on 2012 March 14--17 and May 14--16. [Only the 7(0,7)$-$6(1,6) A$^{++}$ line was separately detected on 2011 February 11.]  The coordinate values of the target, which were used in the observation, were 19$^{h}$33$^{m}$24$^{s}_{\cdot}$249, $+$19$^{\circ}$56$'$55$^{''}_{\cdot}$65 (J2000.0). The receivers used in the present observations are summarized in table~\ref{tab1}. Since the frequency coverage of H40 (2 GHz) is much wider than S40 (0.5 GHz), in some cases we used H40 to cover multiple lines at a time to shorten the total observing time, even though the system temperature of H40 is slightly higher than that of S40. The half-power beam widths (HPBW) of the telescope at $\lambda=$13~mm, 7~mm and 3~mm were roughly 80$^{''}_{\cdot}$0 (at 22 GHz), 39$^{''}_{\cdot}$6 (at 43 GHz) and 16$^{''}_{\cdot}$1 (at 110 GHz), respectively.  

The acousto optical spectrometers, AOS-H and AOS-W, were used, and the details are summarized in table~\ref{tab2}. Since the line intensity of the observed methanol lines was relatively weak, we mainly used AOS-W, of which the velocity resolution is lower than AOS-H. The antenna temperature ($T_{a}$) used in the present paper is corrected for the atmospheric and telescope ohmic loss, but not for the beam nor aperture efficiency. Observations were made in a position-switching mode, and the off-position was chosen to be a location $+$20$'$ away from the object in the right ascension, where contamination is minimized (the background emission at the off-position was carefully examined in our previous study; see, \cite{nak04b}). Data reduction, including flagging bad channels and subtraction of the baseline, was made using the software package NEWSTAR, which has been developed by NRO.

\begin{figure}
 \begin{center}
  \includegraphics[width=12cm]{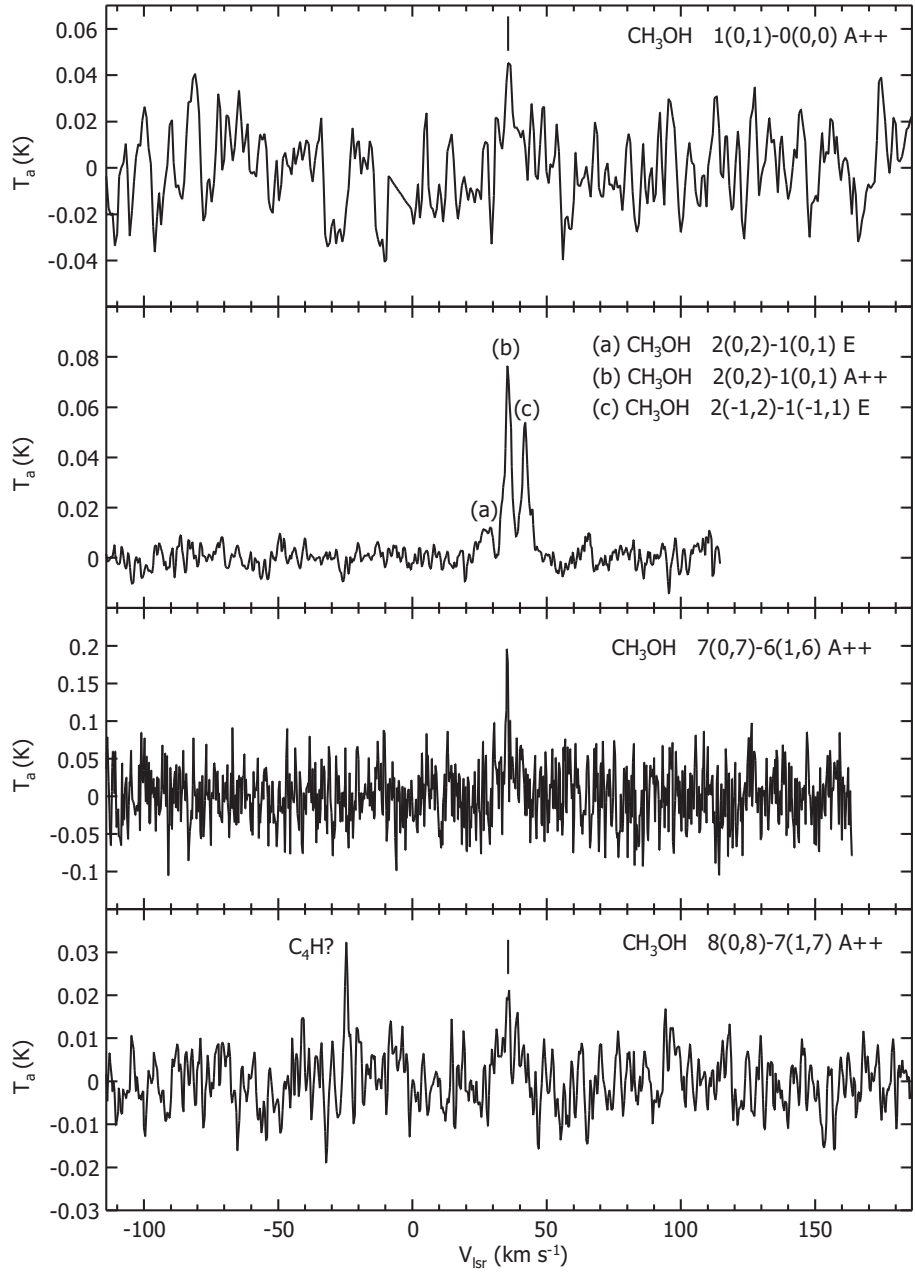} 
 \end{center}
\caption{Spectra of detected CH$_3$OH lines. The velocity for the second top panel is calculated with respect to the 2(0, 2)--1(0, 1) A$^{++}$ line (the middle of the three detected lines). The vertical lines in the top and bottom panels indicates the intensity peaks of the detected lines.}\label{fig1}
\end{figure}

In figure~\ref{fig1}, we present the spectra of 6 detected lines, and the line parameters of the detected lines are summarized in table~\ref{tab3}; the line parameters include rest frequencies, transitions, intensity peak velocities, peak intensity, velocity integrated intensities. The rms noise (in $T_{\rm a}^{*}$) and on-source integration time are also given in table~\ref{tab3}. For non-detections, rms noises (in $T_{\rm a}^{*}$) are are given in table~\ref{tab4}. We use the notation of transitions used in \citet{lov92} throughout the paper. The detected 6 lines include 4 thermal lines (48.37~GHz, 96.739~GHz, 96.741~GHz, and 96.745~GHz) and 2 maser lines (44.07~GHz and 95.17~GHz). The two detected maser lines belong to class I methanol masers (see, e.g., \cite{men91}). As we mentioned above, the 44.07~GHz maser line was separately observed on 2011 February 26, which was roughly 9 months before the observation of the other lines. The integration time and achieved rms noises of the observation on 2011 February 26 is given in table~\ref{tab3}, and those on 2012 May are 940~sec and 0.013~K (i.e., the later observation on 2012 May was more sensitive than the previous one on 2011 February). We observed this maser line again in 2012 May, but the result was negative. In addition, we observed 2 class II methanol lines (i.e., 23.121024~GHz and 108.893929~GHz; \cite{wil84,vor11}), but again the results are negative in the present observation.

All 4 thermal lines exhibit a linewidth of about 3--4~km~s$^{-1}$, while the maser lines show a narrower linewidth of about 1--2~km~s$^{-1}$ (here, the linewidths are measured at the 0-intensity level). The intensity peak velocities of detected lines exhibit consistent values of about 35.5~km~s$^{-1}$ -- 36.1~km~s$^{-1}$ except for the 96.744~GHz line, of which the intensity peak velocity is 39.1~km~s$^{-1}$. This small shift in the velocity is due presumably to the weak intensity of this line (i.e., low signal-to-noise ratio). In figure~\ref{fig1}, the intensities of the 1(0,1)$-$0(0,0) A$^{++}$ and 8(0,8)$-$7(1,7) A$^{++}$ lines are relatively weak compared with other lines (signal-to-noise ratios are roughly 3--4). However, the intensity peak velocities of these lines almost exactly correspond to those of other lines, and additionally the emission is detected in multiple frequency channels; this situation supports that the emission is originated in a natural astrophysical object rather than spurious emission, which is often detected in a single frequency channel. One may think that these could be spurious emission, by chance, lying exactly at the systemic velocity of the object. However, the chance probability of finding such very unusual artificial features in two frequency bands at a time would be negligibly small. Thus, it is natural to conclude that these two lines are the true emission of methanol molecules.

In the bottom panel of figure~\ref{fig1}, we see another possible emission at about $-25$~km~s$^{-1}$. This emission exhibits the intensity of 0.032~K with a signal-to-noise ratio of 4.6~$\sigma$. This emission is close to the C$_4$H 10--9 J=19/2--17/2 line at 95.18894~GHz (the radial velocity of the emission is 36.5 km~s$^{-1}$ if we assume the rest frequency of the C$_4$H 10--9 J=19/2--17/2 line). However, we harbor doubts on this marginal detections for the following two reasons. Firstly, we did not detect another C$_4$H line (10--9 J=21/2--19/2 at 95.15032~GHz), which is within the frequency range of the bottom panel of figure~\ref{fig1}. In fact,  the C$_4$H 10--9 J=21/2--19/2 line was detected toward IRC+10216 with almost the same intensity with the C$_4$H 10--9 J=19/2--17/2 line \citep{gue78}. Secondarily, the linewidth seems to be too narrow as a thermal line; this unusually narrow linewidth is reminiscent of the characteristic of an artificial spurious emission (this is obvious when we compare with the line profiles of other thermal lines; see, Appendix~1, for example). Thus, as the detection is quite marginal, we do not go deeper into scientific discussions about this emission.


\section{Radiative Transfer Modeling of Detected Methanol Lines}
For better understanding of the emission region of methanol lines, we constructed a simple radiative transfer model, which uses the large velocity gradient (LVG) approximation. Dust emission and absorption within the emission region was taken into account in the way described in \citet{sut04}. We assumed that the dust particles are intermixed with gas molecules. The same physical temperature of the gas and dust components is assumed. The molecular emission region was assumed to be spherically symmetric (in the optical sense) and uniform in H$_2$ density, gas and dust temperature, gas-to-dust ratio and methanol fractional abundance. The influence of external infrared sources was not considered. The dust opacity law was chosen as $\tau_{\rm dust} \sim \lambda^{-2}$. We adopted a gas-to-dust mass ratio of 100 and a cross section at 1~mm of $2.6 \times 10^{-25}$~cm$^2$ \citep{she80}.

In addition to the model described in \citet{sut04}, we used collision transition rates based on the model of collisions of methanol molecules with He and para-H$_2$ molecules \citep{cra05} in this work. Recently, newer collisional rates have been published \citep{rab10a,rab10b}, but we would like to note that the difference caused by using the newer rates is negligible for low transition methanol lines, such as the lines analyzed in the present research. The scheme of energy levels in this model includes rotational levels with quantum numbers $J$ up to 22, $|K|$ up to 9; the levels include the rotational levels of the ground, first and second torsionally excited states. In total, 861 levels of A-methanol and 852 levels of E-methanol were considered according to \citet{cra05}.

Through the course of model fitting, we made estimates of the values of hydrogen number density, $n_{\rm H_2}$ [cm$^{-3}$], specific column density of methanol, $N_{\rm CH_3OH} / \Delta V$ [cm$^{-2}$s], gas kinetic temperature, $T_{\rm k}$ [K], using the measured values of ``quasi-thermal'' (i.e. not maser) methanol lines. The brightness temperatures of three lines from quartet lines at 96.741~GHz--96.744~GHz and 48.37~GHz, as well as upper limits for the brightness temperatures of 10 other lines in the frequency range 36.169~GHz--108.893~GHz were considered (see, tables \ref{tab5} and \ref{tab6}). 

In order to estimate these physical parameters, we have searched for a set of parameters which exhibits the best agreement between the values of the calculated brightness temperatures ($T^{\rm mod}_i$) and the measured brightness temperatures ($T^{\rm obs}_i$). This corresponds to finding the minimum of $\chi^2 = \sum_{i}^{N} ((T^{\rm obs}_i - T^{\rm mod}_i )/\sigma_i)^2$, where $\sigma_i$ is the observational uncertainty for a particular line. 

The process was done in two steps. In the first step we explored the parameter space in order to find the approximate location of the $\chi^2$ minimum. For this purpose, we used the database of population numbers for the quantum levels of methanol described in \citet{sal06}. We give a brief description of this database below. 

The kinetic temperature in the database ranges from 10~K to 220~K with a 10~K step. The hydrogen number density varies from 10$^3$ to 10$^9$ cm$^{-3}$ with a step of 0.5~dex. The fractional abundance of methanol molecules relative to molecular hydrogen in the population number database varies from $10^{-9}$ to $10^{-6}$ with a 1~dex step. The lower values of these parameters correspond to the physical conditions of dark molecular clouds, while the higher values of both parameters can occur in shocked molecular material. The outer parts of AGB star envelopes have values of kinetic temperature and hydrogen number density corresponding to the higher ends of the given intervals. The values of the methanol specific column density  ($N_{\rm CH_3OH} / \Delta V$ [cm$^{-2}$s]) in the database range from 10$^8$ cm$^{-2}$ s to 10$^{13}$ cm$^{-2}$ s with a step of 0.25~dex. 

The variable parameter of the source size is introduced in order to take into account beam-dilution effects. According to the results of CO mapping observations \citep{nak04b,nak05}, the source in molecular line emission seems to have a size of the order of 5$''$ -- 15$''$. The analysis made after the first step of the search for the best fit parameters provided the following constraints for the model parameters: 
kinetic temperature $T_{\rm k}$ not higher than 70~K, n$_{\rm H}$ is about 10$^5$ cm$^{-3}$, specific methanol column density is about 10$^8$, and the fractional abundance of methanol is not less than 10$^{-7}$, the source size is about 8.6$''$.

At the second step of the search we made more accurate estimates of the physical parameters with the constraints that were obtained in the first step. We found that within these constraints the fit is not sensitive to fractional abundance values within the interval from 10$^{-7}$ to 10$^{-6}$. For the analysis, we adopted the value for fractional abundance of 10$^{-7}$. We also fixed the value for the source size at 8.6$''$. Since the expected values of the kinetic temperature are rather low ($T_{\rm k}$ not higher than 70~K), we reduced the scheme of the energy levels in order to increase the robustness of the model fit and decrease the computing time. The scheme of methanol energy levels used in the second step was restricted to rotational levels with the quantum numbers $J$ up to 18, $|K|$ up to 8; the scheme included rotational levels of the ground and the first torsionally excited states. In total, 380 levels of A-methanol and 374 levels of E-methanol were considered. 

\begin{figure}
 \begin{center}
  \includegraphics[width=7cm]{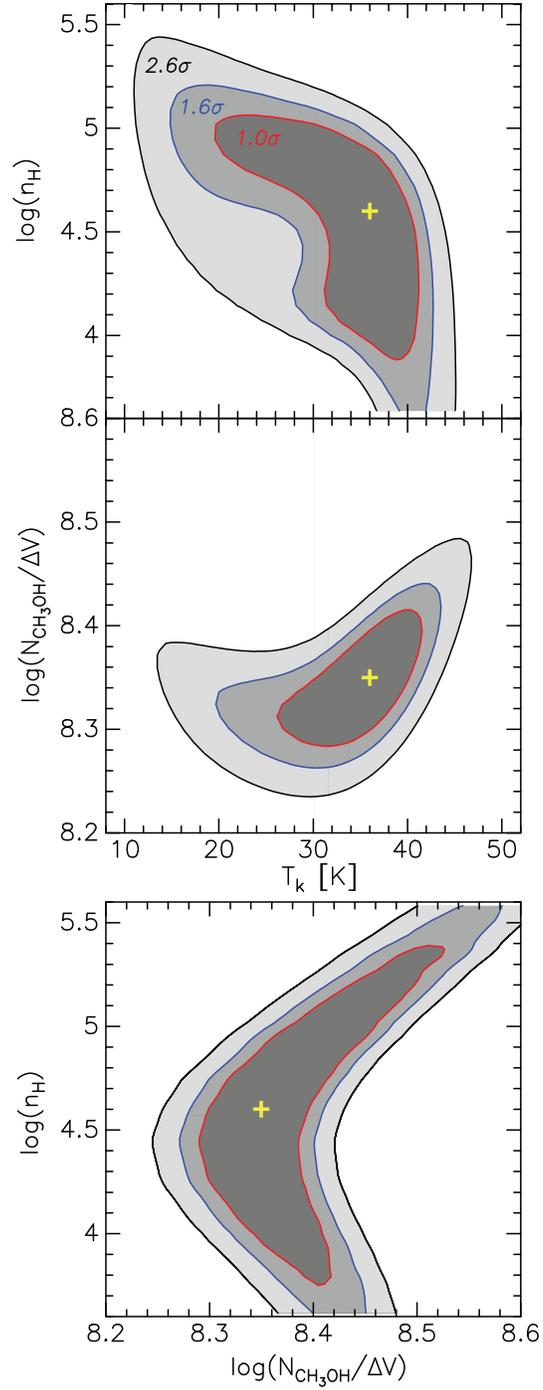} 
 \end{center}
\caption{Visual summary of the model calculation (see, Section 3 for details). The yellow plus-mark represents the best fit solutions.}\label{fig2}
\end{figure}

Consequently, the set of parameters $T_k = 36$~K,  $N_{\rm CH_3OH} /\Delta V = 2.24 \times 10^8$ cm$^{-2}$s ($N_{\rm CH_3OH} = 3.83\times 10^{13}$ cm$^{-2}$), $n_{\rm H_2} =  3.98\times 10^4$ cm$^{-3}$ was found to provide the best fit with $\chi^2 = 0.08$. Upper limits on $T_{\rm br}^{\rm mod}$ for undetected lines are within the error bars [it is noteworthy that the 23.445~GHz, 36.169~GHz, 44.069~GHz, 95.170~GHz lines experience population inversion (maser effect) with this set of parameters; see, table~\ref{tab6}]. To estimate the confidence level of the fit, we used the method given by \citet{lam76}. According to \citet{lam76} for 3 free parameters, a $1\sigma$ confidence interval includes parameters within contours $\chi^2_{\rm min} + 3.5$. Thus, $1\sigma$ confidence intervals are for  $T_k = 20$--$40$~K, $N_{\rm CH_3OH} /\Delta V =$ (1.9--3.3)$\times 10^8$~cm$^{-2}$s, $n_{\rm H_2}  =$  (0.05--2.5)$\times 10^5$~cm$^{-3}$. Estimated physical parameters are typical for molecular clumps experiencing the influence of moderately strong shock waves. The results of the present calculation are visually summarized in figure~\ref{fig2}. In table~\ref{tab6}, the theoretically predicted intensity of the 0(0,0)$-$1($-$1,1) E line (108.8~GHz) is 3.5 times higher than the observational 3~$\sigma$ rms level. This inconsistency is due presumably to the simpleness of the present model. Since the optical depth of this line is significantly larger than other lines, the intensity must be sensitive to the structure of the emission source. In addition, for higher frequency lines, the contribution from dust emission, which is not considered in the present calculation, would be non-negligible, and the dust emission could be stronger in higher frequencies than in lower frequencies. Therefore, to explain all the lines including higher frequency lines, in future we need to consider additional physical conditions, such as the structure of the emission sources and the emission of dust with the dependency on frequencies.


\section{Discussion}
In the present observations, we securely detected methanol lines toward I19312, and derived some informative results: e.g., kinetic temperature, line profiles and detections of class I maser lines, etc.  In the first subsection we compare the sensitivities between the present observations and the previous methanol observations of evolved stars, because many characteristics of I19312 are consistent with those of mass-losing evolved stars. This process is important, because there have been no detections of methanol emission toward evolved stars. We need to clarify the peculiarity of I19312 by comparing the present observation with the previous methanol observations of evolved stars. In the next subsection, we discuss the nature of methanol emission and the possible locations of the emission regions based on the derived results.

\subsection{Comparison of Sensitivities}

As we previously mentioned in Section~1, \citet{cha97} searched the CH$_3$OH 2(0, 2)$-$1(0, 1) A$^{++}$ line at 96.741377~GHz toward 6 evolved stars (i.e., IRC+10420, R Cas, IRC+10011, TX Cam, IK Tau, and OH 231.8+4.2). Since we also observed this line, we shall compare the sensitivities. The rms noise level achieved by \citet{cha97} was typically 0.0028~K. If we assume a conversion factor to Jy is 35.0 $S_\nu$[Jy]/$T_R^*$[K] (see, e.g., \cite{pap12}) for the observation of \citet{cha97}, their typical rms converts to 0.098~Jy. On the other hand, the rms noise level of the present observation in the CH$_3$OH 2(0,2)$-$1(0,1) A$^{++}$ line is 0.009~K ($T_A^*$), and this is converted to 0.036~Jy (the conversion factor to Jy is given in table~\ref{tab3}). This means that the present observation is roughly 3 times more sensitive than \citet{cha97}. However, we should note that even if I19312 was observed in \citet{cha97}, the signal-to-noise ratio could be still more than 3$\sigma$ (strictly, 3.1$\sigma$); therefore, the emission of I19312, could, in principle have been detected by \citet{cha97} if they had observed it.

\citet{cha97} searched for CH$_3$OH 2(0, 2)$-$1(0, 1) A$^{++}$ line emission toward a proto-planetary nebulae, OH 231.8+4.2 (Rotten Egg Nebula), which in our previous work we identified as having some similarities with I19312 \citep{nak00,nak03,nak04a,nak04b,nak05,nak11}. The 12 micron IRAS flux density of OH 231.8+4.2 (18.98~Jy) is similar to that of I19312 (22.47~Jy), and both exhibit very red mid-infrared colors [$\log (F_{25\mu m}/F_{12 \mu m}) > 0.5$]. Unfortunately, however, the sensitivity of \citet{cha97} was exceptionally shallow only for this object (the achieved rms was 0.0043~K in $T_R^*$~[K]). If OH 231.8+4.2 exhibits the methanol line, of which the intensity is equivalent to that of I19312, the emission could be most likely missed due to the expected low signal-to-noise ratio of about 2. Therefore, we note that it is worth to make a more sensitive search toward OH 231.8+4.2 for confirmation. We also note that the distance to OH 231.8+4.2 is somewhat closer to us(1.5 kpc; \cite{cho12}) compared to that to I19312 (3.8 kpc); if the 12 micron IRAS flux is normalized at 3.8~kpc, the flux of OH 231.8+4.2 is roughly 7.6 times weaker than that of I19312.

In addition, some sensitive searches covering methanol lines have been undertaken toward the well-studied carbon star, IRC+10216 \citep{kaw95,cer00,for04,he08}, and the search results were negative in all cases. Even though the characteristics of IRC+10216 are somewhat different from I19312, we shall compare the sensitivities of these observations with the present case, because the methanol searches toward IRC+10216 are the most sensitive search toward evolved stars. 

\citet{kaw95} made a line survey in the 30--50~GHz range using the Nobeyama 45~m telescope; the observed frequency range included the CH$_3$OH 1(0, 1)$-$0(0, 0) A$^{++}$ and 7(0, 7)$-$6(1, 6) A$^{++}$ lines, which are detected in the present observation toward I19312. The rms noise achieved by \citet{kaw95} was 0.008--0.011~K ($T_a^*$). If we observed I19312 at this rms level, the expected signal-to-noise ratio would be 4.1$\sigma$ -- 5.6$\sigma$ for the 1(0, 1)$-$0(0, 0) A$^{++}$ line, and 17.7$\sigma$ -- 24.4$\sigma$ for the 7(0, 7)$-$6(1, 6) A$^{++}$ line. With this high signal-to-noise ratio expected, we can easily understand the peculiarity I19312 in its intensity of methanol emission. We should note that the distance to IRC+10216 is 120--150~pc \citep{gro98,luc99}, while that to I19312 is 3.8~kpc \citep{ima11}: i.e., I19312 is even much further distant than IRC+10216. 

The frequency ranges of \citet{cer00,for04,he08} (129~GHz--267.5~GHz) do not overlap with that of the present observations. However, since our model calculation predicts intensities in higher frequencies, we briefly compare the sensitivities about the high frequency lines. In table~\ref{tab7}, we show a list of the methanol lines that are predicted to exhibit a strong intensity brighter than $T_{\rm br}^{\rm mod} \sim 0.3$~K; these lines are in the frequency range of \citet{cer00,for04} and \citet{he08} except for the two lines at 193~GHz. 

For the 145~GHz and 157~GHz lines [3($-$1, 3)$-$2($-$1, 2) E, 3(0, 3)$-$2(0, 2) A$^{++}$, and 1(0, 1)$-$1($-$1, 1) E], the rms noise levels achieved by the previous observations at 145~GHz and 157~GHz were respectively 0.0135~K and 0.0117~K (\cite{cer00}) and 0.008~K and 0.031~K (\cite{he08}) both in $T_{\rm mb}$. This means that if the model prediction is correct, the 145~GHz and 157~GHz lines should be detected respectively at the signal-to-noise levels of 61$\sigma$ -- 118$\sigma$ and 14$\sigma$ -- 38$\sigma$. For the 241.8~GHz lines [i.e., 5($-$1, 5)$-$4($-$1, 4) E, and 5(0, 5)$-$4(0, 4) A$^{+}$], the rms noise levels achieved by the previous observations were 0.002~K (\cite{for04}) and 0.015~K (\cite{he08}) both in $T_{\rm mb}$. Therefore, the expected signal-to-noise level is 22$\sigma$ -- 198$\sigma$ for this line. 

As described above, we have to say that the properties of methanol lines in I19312 are far different from those expected in evolved stars. If I19312 is really an evolved star, the hidden mechanism, which causes the peculiarly strong methanol emission, is likely not directly related to the nature of the star, but rather the ambient materials and/or the interaction with ambient material presumably play a role.

\subsection{Characteristics of Methanol Emission}

\begin{figure}
 \begin{center}
  \includegraphics[width=8cm]{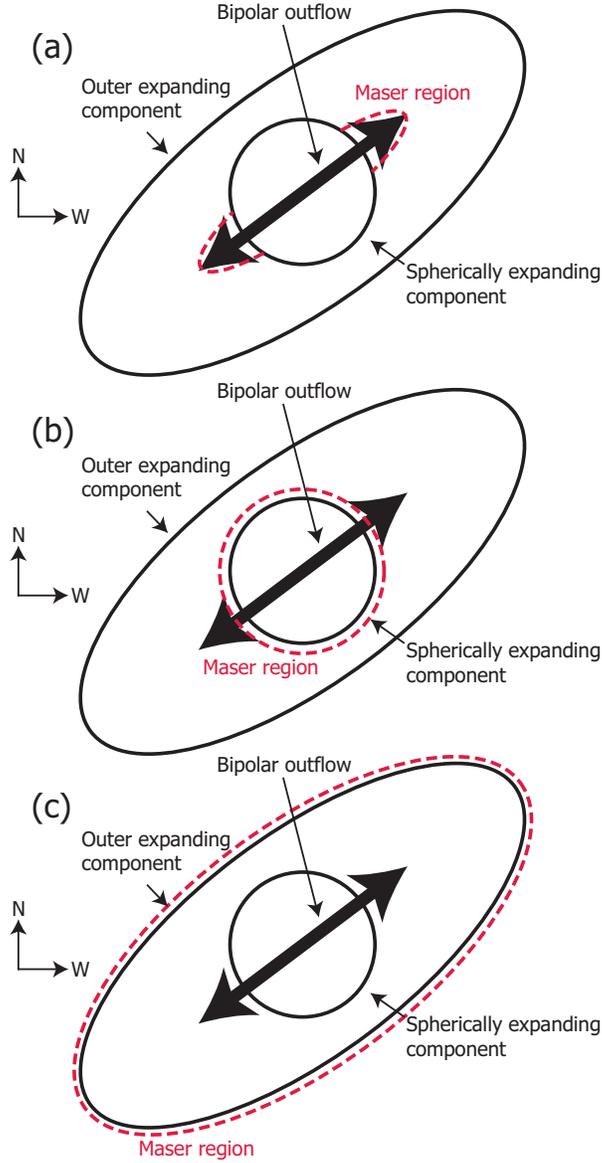} 
 \end{center}
\caption{Schematic view of the molecular components of IRAS~19312+1950. The panels (a), (b) and (c) indicate different possible emission regions of class I methanol masers; the red broken lines represent the possible emission regions, in which shocked gas is generated by hydrodynamical interaction between different kinematic components. The relative sizes between different components is arbitrary; the actual sizes of the spherically expanding component and the outer expanding component (in the long axis) are roughly 10$''$ and 20$''$, respectively (see, \cite{nak05}). The actual size of the bipolar outflow is unknown, because so far it has been observed only in maser lines \citep{nak11}. The position angle of the jet axis is known to be roughly $-37^{\circ}$.}\label{fig3}
\end{figure}

The properties of detected methanol thermal lines are not consistent with those of a spherically expanding component, which is lying at the center of the nebulosity (\cite{nak05,mur07}, see, figure~\ref{fig3}). There are two pieces of evidence supporting this statement. (1) The line profiles of the methanol thermal lines are not consistent with the kinematics of the spherically expanding component. Our previous radio observations of I19312 \citep{nak04b,nak05} revealed that the spherically expanding molecular component exhibits an expanding velocity of about 40~km~s$^{-1}$. This expanding velocity is clearly inconsistent with the narrow linewidth of the detected methanol thermal lines showing a linewidth of 3--4~km~s$^{-1}$. (2) The derived kinetic temperature of methanol lines (20--40~K) is most likely not consistent with the temperature of the spherically expanding component. This is because the dust temperature of the spherically expanding component is estimated to be 150~K--200~K using the mid-infrared photometric data \citep{nak04b}. The high dust temperature presumably suggests that the spherically expanding component shows a much higher kinetic temperature than that measured using methanol lines. 

Based on the estimated kinetic temperature, we can roughly infer the location of the methanol emission region. The low temperature naturally suggests that the emission region should be at the outer region of the nebulosity, which is outer than the spherically expanding component. Our previous radio interferometric observation revealed that the spherically expanding component is surrounded by an outer expanding component with a size of about 15$''$--20$''$ (see, figures 7 and 8 in \cite{nak05}; see, also figure~\ref{fig3}); the outer expanding component exhibits a bipolar structure. The expanding velocity of the outer expanding component is about 3--5~km~s$^{-1}$, which is consistent with the linewidth of the methanol thermal lines. The kinetic temperatures of the outer expanding component, which are determined by the intensities of CO and NH$_3$ lines, are 13~K and 19~K, respectively \citep{deg04}; those are very close to the lowest possible kinetic temperature obtained from methanol lines. Therefore, the outer expanding component seems to be one of the possible sites of methanol molecules. 

The detection of class I methanol maser lines gives us a further hint about the site of methanol molecules. It is well known that there are two classes in methanol masers (i.e., class I and II), and the emission region of the two classes of masers (and therefore excitation mechanisms) are different each other (see, e.g., \cite{men91}). Class I masers are predominantly detected in shocked molecular gas \citep{kur04}, while class II masers are caused by radiation pumping. In the case of star forming regions, class I methanol masers are often detected along the periphery of a bipolar outflow, which induces shocked molecular gas \citep{vor06}. According to these properties of methanol maser, there possibly exists a shocked region (or regions) in the system or I19312. 

We would like to note that even small relative velocity (less than 10~km~s$^{-1}$ ) between two colliding components can cause class I masers \citep{vor06}. In fact, in general, class I methanol maser emission is observed very close to the systemic velocity of the source. This suggests that in most cases the shocks are low velocity and/or primarily perpendicular to the line of sight. This fact suggests that there are three possible sites of class I methanol masers within the system of I19312. The first possible site of class I maser is, of course, the periphery of a bipolar flow \citep{nak11}. Even though the spherically expanding component does not seem to include methanol molecules as discussed above, the bipolar flow can generate masers if the tip of the jet goes beyond the outer edge of the spherically expanding component [see {\bf (a)} in figure~\ref{fig3}].  Then, the periphery of the spherically expanding component also may be a possible emission region of class I methanol masers [see {\bf (b)} in figure~\ref{fig3}], because the expanding velocity of the spherically expanding component is 20--30~km~s$^{-1}$ \citep{nak05}, which seems to be sufficient enough to induce class I masers by a shock interaction. 

Another possible site of class I masers may be the periphery of the outer expanding component [see {\bf (c)} in figure~\ref{fig3}], because we recently detected ambient molecular gas outside of the outer expanding component by an on-the-fly (OTF) mapping in the CO $J=1$--0 line (Nakashima et al., in preparation).  Even though the expanding velocity of the outer expanding component is relatively small (3--5~km~s$^{-1}$), shocked molecular gas still may be induced if the gas motion in the perpendicular direction is assumed. Interestingly, near-infrared images taken by the Subaru 8~m telescope and the University of Hawaii 2.2~m telescopes revealed that the infrared nebulosity shows a cometary tail extended in the south-west direction \citep{deg04,mur07}; this may suggest that I19312 may have a relative velocity against to the ambient molecular gas along this direction.  

The true origin of the outer expanding component of I19312 would depend on the location of the emission site of methanol masers. If the cases {\bf (a)} and {\bf (b)} are the case, methanol molecules must be preexisting in the outer expanding component. In such a case, the origin of the outer expanding component must different from that of the spherically expanding component and the bipolar jet, because it is hard to explain the existence of methanol molecules in a stellar envelope of a mass-losing evolved star. On the other hand, if class I maser is induced in the periphery of the outer expanding component (and methanol molecules are not included in the outer expanding component), the outer expanding component is still possibly explained with a common origin with the spherically expanding component and bipolar jet, at least, in terms of the methanol chemistry. Thus, a high angular-resolution radio interferometry, which enables us to specify the location of the emission regions of class I maser, would be helpful to constrain further the interpretations of I19312.

Until now, class I methanol masers have been detected toward high-mass star formation regions (see, e.g. \cite{kur04}), lower-mass star formation regions (see, e.g. \cite{kal10}), interaction between supernova remnant and molecular clouds (see, e.g. \cite{pih11}) and starburst galaxies (see, e.g. \cite{ell14}).  Even though there have been no conclusive evidences about the origin of I19312 so far, many observational properties of I19312 are consistent with those of an a mass-losing evolved star showing a bipolar flow (see, e.g., \cite{nak11}). If the central star of I19312 is an evolved star, the class I maser toward I19312 should be classified into a new category of class I methanol masers, which is caused by an interaction between the outflow of an evolved star and ambient materials.


\section{Summary}
We have reported the results of a systematic methanol observation of the enigmatic IRAS object 19312+1950. We detected 6 methanol lines including 4 thermal lines at 48.372~GHz, 96.739~GHz, 96.741~GHz, and 96.745~GHz, and 2 maser lines at 44.069~GHz and 95.170~GHz. The detected maser lines belong to the category of class I masers, which are known as a tracer of shocked molecular gas. For better understanding of the physical conditions of the emission region, we constructed a simple radiative transfer model, and derived some physical parameters including kinetic temperature and relative abundance. The line profiles and derived excitation temperature suggest that the methanol emission region does not coincide with the spherically expanding molecular outflow lying at the center of the I19312 system. The detections of class I methanol masers suggest that there is a shocked gas in the system of I19312. We gave an implication about the location of possible methanol emission regions based on the present results. If the central star of I19312 is an evolved star, as suggested in the past, the class I maser detected in the present observation is the first case detected in an interaction between evolved star outflow and ambient molecular gas.


\bigskip
This work is supported by a grant awarded to JN from the Research Grants Council of Hong Kong (project code: HKU 703308P; HKU 704209P; HKU 704710P). AMS was supported by the Russian Science Foundation (grant number 15-12-10017). SVS was supported by the Ministry of Education and Science of the Russian Federation (state task No. 3.1781.2014/K). YZ also thanks the Hong Kong General Research Fund (HKU 7027/11P and HKU 7062/13P) for the financial support of this study. The model analysis and a part of paper writing have been done while JN was staying at the Department of Astronomy and Geodesy in the Ural Federal University in November and December 2013. Nobeyama Radio Observatory is a branch of the National Astronomical Observatory of Japan, National Institutes of Natural Sciences. A part of the data was acquired within a telescope time allocated for a project of Daniel Tafoya.


\appendix
\section{Other Thermal Lines Detected in the Present Observation}
In the present observations, we detected 3 thermal lines other than methanol lines. In table~\ref{taba1}, the line parameters of the detected 3 thermal lines are summarized, and in figure~\ref{figa1}, the spectra of the lines are given. Among the 3 lines, the SO N, $J=3$, 2--2, 1 line at 109.252212~GHz and the HC$_3$N 12--11 line at 109.173638~GHz were already detected by \citet{deg04}, and they discussed chemistry of the molecular component by constructing a line radiative transfer model. The HC$_3$N 4--3 $F=5$--4 line at 36.392365~GHz is newly detected in the present observation. The linewidth of the SO N, $J=3$, 2--2, 1 line ($\sim$9--10km~s$^{-1}$) is slightly wider than that of the HC$_3$N lines ($\sim$6--7km~s$^{-1}$).



\newpage
\begin{table}
\caption{Receivers Used in the Observation}\label{tab1}
\begin{center}
\begin{tabular}{lcc}
\hline \hline
Freq.~range & Receiver & System temp.\\
(GHz) &  & (K) \\
\hline 
20.0--25.0  &  H22 (HEMT)  & 120--180\\
35.0--50.0  &  S40 (SIS)   & 180--230\\ 
42.0--44.0  &  H40 (HEMT)  & 200--250\\
80.0--116.0 &  T100V/H (SIS)    & 120--250\\ 
\hline
\end{tabular}
\end{center}
\end{table}

\newpage
\begin{table}
\caption{Spectrometers Used in the Observation}\label{tab2}
\begin{center}
\begin{tabular}{lcc}
\hline \hline
 & AOS-W & AOS-H\\
\hline 
Frequency channel  &  2048  & 2048\\
Frequency coverage  &  250~MHz  & 40~MHz\\ 
Vel. resolution at 22 GHz  &  3.4 km~s$^{-1}$ & 0.5 km~s$^{-1}$\\
Vel. resolution at 43 GHz  &  1.7 km~s$^{-1}$ & 0.3 km~s$^{-1}$\\ 
Vel. resolution at 43 GHz  &  0.7 km~s$^{-1}$ & 0.1 km~s$^{-1}$\\
\hline
\end{tabular}
\end{center}
\end{table}

\newpage
\renewcommand{\thefootnote}{\fnsymbol{footnote}}
\begin{table}
\caption{Parameters of Detected CH$_3$OH Lines (temperatures are in $T_{\rm a}^{*}$).}\label{tab3}
\begin{center}
\begin{tabular}{llccccc}
\hline \hline
Rest freq. & Transision & $V_{\rm peak}$ & $I_{\rm peak}$ & $I_{\rm int}$ & rms & Integ. time \\
(GHz) &  & (km~s$^{-1}$) & (K) & (K~km~s$^{-1}$) & (K) & (sec)\\
\hline 
44.069476$^{*\dagger}$ & 7(0,7)$-$6(1,6) A$^{++}$ &  36.1 & 0.195 & 0.560 & 0.037 & \phantom{0}560\\
48.372456$^{\dagger}$ & 1(0,1)$-$0(0,0) A$^{++}$ &  35.8 & 0.045 & 0.249 & 0.018 & 2540\\
95.169516$^{\ddagger}$ & 8(0,8)$-$7(1,7) A$^{++}$ & 36.0 & 0.021 & 0.115 & 0.007 & 3040\\
96.739363$^{\ddagger}$ & 2($-$1,2)$-$1($-$1,1) E & 35.8 & 0.054 & 0.205 & 0.009 & 4300\\
96.741377$^{\ddagger}$ & 2(0,2)$-$1(0,1) A$^{++}$ & 35.5 & 0.076 & 0.228 & 0.009 & 4300\\
96.744549$^{\ddagger}$ & 2(0,2)$-$1(0,1) E & 39.1 & 0.012 & 0.064 & 0.009 & 4300\\
\hline
\multicolumn{7}{l}{
\footnote{}{Separately observed on 2011 February 26.}
}\\
\multicolumn{7}{l}{
\footnote{}{Conversion factor to Jy at these frequencies is 2.89 Jy/K.}
}\\
\multicolumn{7}{l}{
\footnote{}{Conversion factor to Jy at these frequencies is 4.00 Jy/K.}
}
\end{tabular}
\end{center}
\end{table}

\newpage
\begin{table}
\caption{Noise levels and Integration Times of Non-Detections (temperatures are in $T_{\rm a}^{*}$)}\label{tab4}
\begin{center}
\begin{tabular}{llcc}
\hline \hline
Rest freq. & Transition & rms & Integ.~time\\
(GHz) &  & (K) & (sec)\\
\hline 
\phantom{0}23.121024  &  9(2,7)$-$10(1,10) A$^{++}$         & 0.010 & 3200\\
\phantom{0}23.444778  &  10(1,9)$-$9(2,8) A$^{--}$          & 0.010 & 3200\\ 
\phantom{0}36.169290  &  4($-$1,4)$-$3(0,3) E                & 0.013 & 1280\\
\phantom{0}44.069476  &  7(0,7)$-$6(1,6) A$^{++}$            & 0.013 & \phantom{0}940\\ 
\phantom{0}48.247572  &  1(0,1)$-$0(0,0) E  $\nu_t=1$        & 0.022 & 2540\\  
\phantom{0}48.376889  &  1(0,1)$-$0(0,0) E                   & 0.015 & 2540\\  
\phantom{0}94.541806  &  8(3,5)$-$9(2,7) E                   & 0.006 & 3040\\  
\phantom{0}95.169516  &  8(0,8)$-$7(1,7) A$^{++}$            & 0.006 & 3040\\ 
\phantom{0}96.755507  &  2(1,1)$-$1(1,0) E                   & 0.004 & 4300\\   
108.893929  &  0(0,0)$-$1($-$1,1) E                & 0.010 & 4300\\   
\hline
\end{tabular}
\end{center}
\end{table}

\newpage
\begin{table}
\caption{Detected lines taken into consideration in the LVG calculation.}\label{tab5}
\begin{center}
\begin{tabular}{llccc}
\hline \hline
Frequency & Transition & $T_{\rm mb}^{\rm mod}$ & $T_{\rm mb}^{\rm obs}$ (rms) & $\tau$ \\
(GHz) &  & (K) & (K) &  \\
\hline 
48.372456  &    1(0,1)$-$0(0,0) A$^{++}$    & 0.06  & 0.06 (0.02)  & $1.65\times 10^{-2}$  \\
96.739363  &    2($-$1,2)$-$1($-$1,1) E     & 0.13  & 0.13 (0.01)  & $6.91\times 10^{-2}$  \\   
96.741377  &    2(0,2)$-$1(0,1) A$^{++}$    & 0.18  & 0.18 (0.01)  & $7.12\times 10^{-2}$  \\
96.744549  &    2(0,2)$-$1(0,1) E           & 0.03  & 0.03 (0.01)  & $4.70\times 10^{-2}$  \\   
\hline
\end{tabular}
\end{center}
\end{table}

\newpage
\begin{table}
\caption{Lines used only for upper limit confirmation.}\label{tab6}
\begin{center}
\begin{tabular}{llccc}
\hline \hline
Frequency & Transition & $T_{\rm mb}^{\rm mod}$ & $T_{\rm mb}^{\rm obs}$ (Upper limit) & $\tau$ \\
(GHz) &  & (K) & (K) &  \\
\hline 
\phantom{0}23.121024  &  9(2,7)$-$10(1,10) A$^{++}$         & 0.00    & 0.06  &  $\phantom{0}1.1\times 10^{-6}$\\
\phantom{0}23.444778  &  10(1,9)$-$9(2,8) A$^{--}$          & 0.00    & 0.05  &  $-8.7\times 10^{-6}$\\ 
\phantom{0}36.169290  &  4($-$1,4)$-$3(0,3) E                & 0.04   & 0.09  &  $ -3.9\times 10^{-2}$\\
\phantom{0}44.069476  &  7(0,7)$-$6(1,6) A$^{++}$            & 0.00    & 0.08  &  $ -3.2\times 10^{-3}$\\ 
\phantom{0}48.247572  &  1(0,1)$-$0(0,0) E  $\nu_t=1$        & 0.00    & 0.06  &  $  \phantom{0}2.6\times 10^{-5}$\\  
\phantom{0}48.376889  &  1(0,1)$-$0(0,0) E                   & 0.01    & 0.06  &  $  \phantom{0}2.0\times 10^{-2}$\\  
\phantom{0}94.541806  &  8(3,5)$-$9(2,7) E                   & 0.00    & 0.06  &  $  \phantom{0}1.1\times 10^{-4}$\\  
\phantom{0}95.169516  &  8(0,8)$-$7(1,7) A$^{++}$            & 0.00    & 0.06  &  $ -8.1\times 10^{-4}$\\ 
\phantom{0}96.755507  &  2(1,1)$-$1(1,0) E                   & 0.01    & 0.01  & $  \phantom{0}1.1\times 10^{-2}$\\   
108.893929  &  0(0,0)$-$1($-$1,1) E                          & 0.07    & 0.02  &  $ \phantom{0}9.1\times 10^{-2}$\\   
\hline
\end{tabular}
\end{center}
\end{table}

\newpage
\begin{table}
\caption{Bright Lines ($T_{\rm br}^{\rm mod}>0.3$~K) expected from the model.}\label{tab7}
\begin{center}
\begin{tabular}{llc}
\hline \hline
Frequency & Transition & $T_{\rm br}^{\rm mod}$  \\
(GHz) &  & (K)  \\
\hline 
145.097443  & 3($-$1,3)$-$2($-$1,2) E         & 0.827   \\
145.103194  & 3(0,3)$-$2(0,2) A$^{++}$          & 0.945   \\
157.270818  & 1(0,1)$-$1($-$1,1) E          & 0.441   \\
193.441610  & 4($-$1,4)$-$3($-$1,3) E          & 0.633   \\
193.454370  & 4(0,4)$-$3(0,3) A$^{++}$          & 0.673   \\
241.767247  & 5($-$1,5)$-$4($-$1,4) E          & 0.333   \\
241.791367  & 5(0,5)$-$4(0,4) A$^{+}$          & 0.354   \\ 
\hline
\end{tabular}
\end{center}
\end{table}

\setcounter{table}{0}
\renewcommand{\thetable}{A--\arabic{table}}

\newpage
\begin{table}
\caption{Parameters of Detected LinesEother than methanol (temperatures are in $T_{\rm a}^{*}$).}\label{taba1}
\begin{center}
\begin{tabular}{lllccccc}
\hline \hline
Molecule & Rest freq. & Transision & $V_{\rm peak}$ & $I_{\rm peak}$ & $I_{\rm int}$ & rms & Integ. time \\
 & (GHz) &  & (km~s$^{-1}$) & (K) & (K~km~s$^{-1}$) & (K) & (sec)\\
\hline 
SO & 109.252212 & N,$J=3$,2--2,1 & 36.3 & 0.060 & 0.385 & 0.007 & 4300\\
HC$_3$N & \phantom{0}36.392365 & 4--3 $F=5$--4 & 36.8 & 0.080 & 0.301 & 0.014 & 1280\\
HC$_3$N & 109.173638 & 12--11 & 36.5 & 0.095 & 0.274 & 0.007 & 4300\\
\hline
\end{tabular}
\end{center}
\end{table}


\newpage

\newpage

\newpage

\setcounter{figure}{0}
\renewcommand{\thefigure}{A--\arabic{figure}}

\newpage
\begin{figure}
 \begin{center}
  \includegraphics[width=12cm]{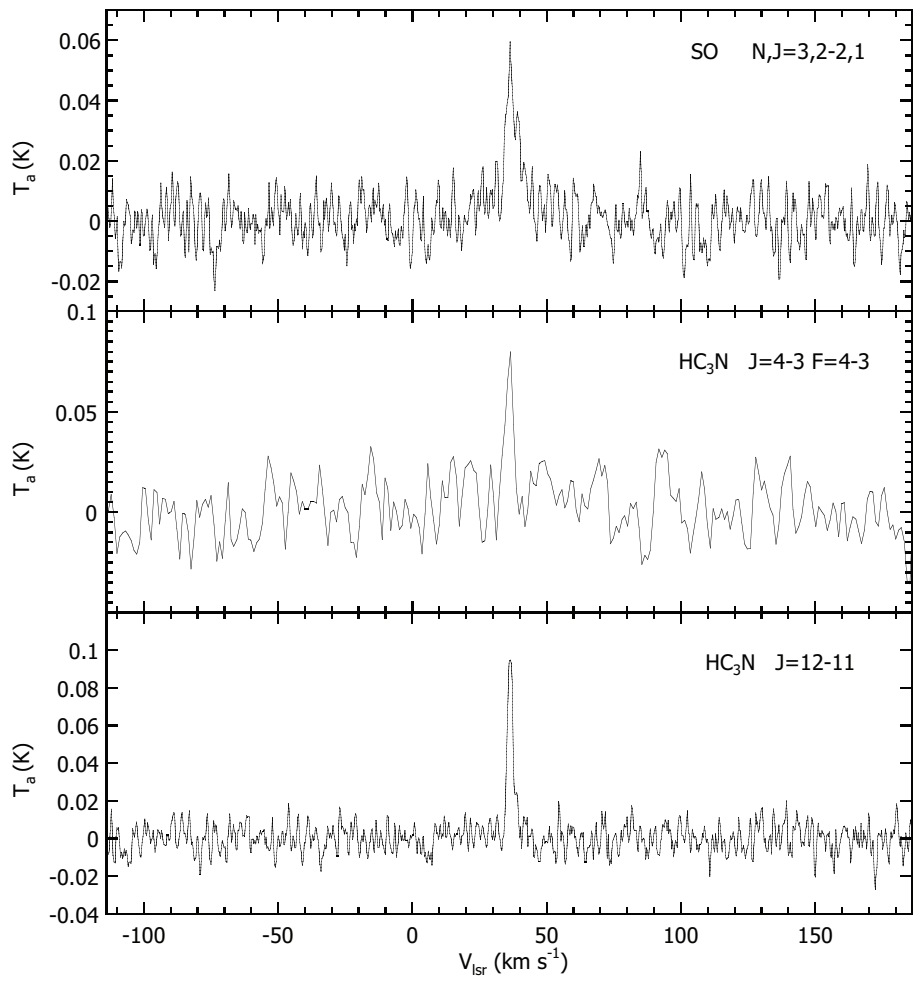} 
 \end{center}
\caption{Spectra of detected thermal lines, which are other than methanol lines.}\label{figa1}
\end{figure}

\end{document}